\documentclass[a4paper]{article}

\usepackage{INTERSPEECH2022}

\usepackage{multirow}
\usepackage{import}

\usepackage{arydshln}

\newcommand\blfootnote[1]{%
  \begingroup
  \renewcommand\thefootnote{}\footnote{#1}%
  \addtocounter{footnote}{-1}%
  \endgroup
}

\title{Comparison and Analysis of New Curriculum Criteria for End-to-End ASR}
  
\name{Georgios Karakasidis, Tam\'as Gr\'osz, Mikko Kurimo}
\address{ Department of Signal Processing and Acoustics, Aalto University, Finland }
\email{\{firstname.lastname\}@aalto.fi}

\begin{document}
\maketitle
\begin{abstract}
It is common knowledge that the quantity and quality of the training data play a significant role in the creation of a good machine learning model. In this paper, we take it one step further and demonstrate that the way the training examples are arranged is also of crucial importance. Curriculum Learning is built on the observation that organized and structured assimilation of knowledge has the ability to enable faster training and better comprehension. When humans learn to speak, they first try to utter basic phones and then gradually move towards more complex structures such as words and sentences. This methodology is known as Curriculum Learning, and we employ it in the context of Automatic Speech Recognition. We hypothesize that end-to-end models can achieve better performance when provided with an organized training set consisting of examples that exhibit an increasing level of difficulty (i.e. a curriculum). To impose structure on the training set and to define the notion of an easy example, we explored multiple scoring functions that either use feedback from an external neural network or incorporate feedback from the model itself. Empirical results show that with different curriculums we can balance the training times and the network's performance.
\end{abstract}

\section{Introduction}

In recent years, we saw a considerable shift towards unified, so-called end-to-end (e2e) models in the automatic speech recognition (ASR) field~\cite{Sklyar2021StreamingMA,Li2021ABA,Mahadeokar21FT}. These new architectures are based on the philosophy that a joint model should be used instead of multiple separate components. Unfortunately, the use of these compact new models is not straightforward. They require special training methods and extreme care, as they must perform a very complex task without the aid of other models. 
Perhaps this complexity is why there is still a gap between the performances of the new e2e and traditional hybrid models, especially when the amount of training data is limited. In this work, we attempt to facilitate the learning process by testing various curriculums.

Curriculum learning (CL) for deep learning was introduced and popularized by~\cite{bengio2009curriculum}, but it has a long history, reaching all the way back to~\cite{ELMAN1993CL}. It is inspired by the way humans learn and proposes that we apply the same \textit{starting small} concept to machine learning models. Hypothetically, employing the right curriculum has the potential to speed up convergence and lead to a more accurate model~\cite{bengio2009curriculum}.

Although curriculum learning has been around for years, not all fields have adopted it yet. It is widely used in reinforcement learning~\cite{Narvekar2020CurriculumLF}, but within speech recognition, it only had a limited usage. 
In speech recognition systems, especially amidst e2e models, a relatively simple curriculum criterion gained fame; sorting the utterances based on their duration. This approach quickly became the standard practice adopted by many systems~\cite{amodei2016deep}.
In~\cite{braun2017curriculum}, we can see a signal-to-noise ratio-based (SNR) curriculum system, while~\cite{RANJAN2021123} suggests that the distance from the microphone can be used to determine a good curriculum.  
 An alternative was proposed by~\cite{zeyer18interspeech}, which starts the training with a reduced set of labels before switching to the character-based labels.
Lastly, an interesting recent work~\cite{Pratap2020} employs CL  to add more languages to a multilingual system incrementally. Without it, the authors faced convergence issues when all languages were taught simultaneously.

Looking at the already published works, we can see that CL has potential in ASR, but only a few simple curriculum criteria have been tested so far. Here, we argue that the standard practice of assuming that longer utterances are harder to learn is not the best solution. We propose new ways of calculating the difficulty from the loss or the recognition results and showcase that these new criteria are superior to the standard length-based one. 




\section{Curriculum Learning}

In the context of machine learning, CL aims to rank the training examples based on their difficulty. Each example is assigned a score and then the data set is ordered accordingly (from the easiest to the hardest example) with the goal of assisting the learning process. Empirical results show that this approach, when done properly,  can speed up convergence and improve the stability of the training process of neural networks~\cite{kim2018improved}.

We can think of CL as a way to provide guidance to a training model. A commonly used analogy to the real world, is that of the teacher-student relationship. The teacher has to create the curriculum in a way so that the student will neither get bored of the easy material, nor get overwhelmed by the hard examples. Instead, there should be an increasing level of difficulty throughout the curriculum. Note that this has nothing to do with the teacher-student transfer learning technique that is commonly used in deep learning.

To create an efficient CL algorithm, we need a suitable scoring function, capable of estimating the difficulty of each training sample. The most common approaches for creating such functions can be grouped into the following three categories:

The \textbf{metadata-based approach} estimates the difficulty scores based on some meta-information (e.g. utterance duration or SNR) at the beginning of the training process. This is a \textit{static} solution since the curriculum is established before training, and the order is kept fixed.

The \textbf{transfer-learning approach} relies on an external, already trained teacher model that is used to infer the training data before the first epoch \cite{hacohen2019power}. The assumption is that the external model should recognize the easy samples with fewer errors than the difficult ones. This method utilizes the outputs of the teacher to sort the utterances at the start of the training, so this is a static approach, as well.

The \textbf{adaptive approach} proposes to sort the examples adaptively using feedback from the student neural network we are training. This approach addresses the fact that the difficulty of the examples (as perceived by the model) could change as the training progresses. One can view this approach as dynamically adjusting the curriculum to the actual state of the model.

To continue the analogy of CL with human learning, the \textit{transfer-learning} approach is equivalent to a teacher network trying to help a student, while in the \textit{adaptive} approach the student is an autodidact trying to adapt to the learning difficulty. Finally, the \textit{metadata} approach consists of a non-adaptive teacher that simply follows guidelines, and there is no knowledge transmission.

In this work, we plan to compare the adaptive and transfer-learning based solutions, to the widely used metadata-based approach, which (in our case) uses the duration of the utterance as an indicator of its difficulty ~\cite{amodei2016deep}. Beyond that, we will also test the performance of models trained with randomly shuffled data.

\subsection{Scoring Function}
\label{scoring_functions}

Commonly, neural networks that use a variant of stochastic gradient descent for their weight re-parameterization, randomly shuffle the data before training. When curriculum learning is applied, the data points are not randomly arranged but rather ordered. Let us denote our training set as $\mathbb{X}$ and define $f: \mathbb{X} \rightarrow \mathbb{R}$ as a \textit{scoring function}. The consequent ordering of the examples will happen in so that $i<j$ for data points $x_i$ and $x_j$ (i.e. the $i$-th element is considered easier than the $j$-th one) if $f(x_i) < f(x_j)$. Choosing a good scoring function is the primary task CL tries to solve. $f$ has to meaningfully encode the acquired knowledge with respect to the notion of difficulty \cite{hacohen2019power}.

To test our hypothesis regarding the importance of updating the CL scores during training, we have experimented with a range of scoring functions, which are discussed below.

\textbf{Duration-based}: A metadata-based approach where the training examples are sorted based on their duration (a shorter duration implies an \textit{easier} example). This approach is the current state-of-the-art solution \cite{watanabe2018espnet}.

\textbf{Loss-based}: We store the loss values of each utterance and at the end of each epoch we re-order the examples based on these values. The rationale behind this solution is that high loss values mean that the model needs to change its parameters more than in the case of lower loss values. Thus we assume that higher loss corresponds to greater difficulty. Examples with the same loss values are sorted with respect to their duration.

\textbf{Metric-based}: After calculating the loss, a decoding step can be performed and a score can be assigned to the corresponding data points based on the word error rates (WER) produced by the model under training. Using the WER directly to assess the difficulty has the advantage that CL now relies on the metric used in the final evaluation of the models. Similar to the loss-based solution, the training examples with the same scores are ordered based on a combination of the prediction confidences and the durations. We assume that high WER values imply harder examples.

\textbf{Loss/Metric-based with uniform Mixing}: A regularized version of the previous two methods. In particular, after acquiring the ordered training set, we cut it into three parts (easy, medium and hard level examples). We then equally mix the easy examples with some hard and medium-level ones. In this work, we have chosen to mix $20\%$ of the easy examples with medium and hard-level ones (due to empirical results). Further implementation notes can be found in the project's github repository. A similar method was employed in \cite{zaremba-suts-learning-lstms} where the authors found that exposing the model to a few difficult examples can improve performance. We argue that this mixing procedure aids the CL system to avoid overfitting, which is a real risk at the start of each epoch, when only easy examples are shown.

To the best of our knowledge, these methods have not been explicitly utilized in the field of ASR before. Most CL-related research in ASR has focused on the metadata approach using mainly the duration~\cite{amodei2016deep,ravanelli2021speechbrain} and signal-to-noise ratio statistics~\cite{braun2017curriculum}. The main shortcoming of these approaches compared to our solutions is their simplicity, since metadata are not always good indicators of the utterances' difficulties. They completely ignore the changes happening in the model during training, which would warrant a re-evaluation of the difficulties of the examples, similar to how teachers adapt their curriculum.

    
    

\subsection{Pacing Function}
\label{pacing_func}

Another way to transform the curriculum is to alter the pace by which the data are presented to the network. A function $g$ is called a pacing function if it transforms a training set $\mathbb{X}$ to a sequence of training subsets $\left[ \mathbb{X}_1, ..., \mathbb{X}_N\right]$ where $N$ is the total number of epochs the model is going to train. Each subset $\mathbb{X}_i$ is going to be used to train the model at epoch $i$. In general, the size of each $\mathbb{X}_i$ can either remain stable (i.e the model will be trained on multiple subsets of the same size), or increase. Here we will only focus on the latter and start by sampling a small percentage of our data, which will over-time increase to the full set of training examples.

The way each subset is acquired by $g$ is through random sampling from $\mathbb{X}$ and the consequent application of a scoring function. In our case, $g$ is applied to our training set every $M$ epochs ($1 \leq  M < N$), i.e. at epochs $i=1, M, 2M, ..., N$ (is always applied at last epoch, even if $M$ is not divisible by $N$). A prerequisite is that the model will see the whole training data in at least one epoch. 

Pacing functions offer shorter training times (by using less data during the first epochs) and a smoother transition from easy to difficult examples. In our work, we experiment with a single pacing function that at each epoch $i=1, M, ..., N$, samples $p_i\%$ of the training data, given by $p_i = p_0 \times \delta^{i/step}$,
where $i$ is the current epoch, $p_0$ is the initial percentage, $\delta$ is the increase factor, and $step$ is a normalizer that prevents huge steps.


\section{Data}

To determine whether CL indeed helps the training process of e2e ASR systems, we tested a range of different setups. All of our results are based on the \textit{Lahjoita puhetta} (\textit{Donate speech}) corpus~\cite{Moisio2022LP}, which is a collection of colloquial Finnish speech. The dataset consists of over twenty thousand speakers from different regions of Finland and a variety of age groups \cite{Moisio2022LP}. Recording was performed by the users' own devices, resulting in a range of different noise levels in the data.

For our experiments, we used a transcribed subset of $1601.5$ hours for the training set, $10.5$ hours for the validation set, and $10.4$ hours for the test set. Initially, the audio data were not segmented and there were utterances of different lengths. This would result in excessive padding and would prevent us from using a large batch size.  To make the data suitable for e2e models, we segmented all long utterances into 10-second chunks, using the best hybrid ASR system described in~\cite{Moisio2022LP} to detect the silences where the cuts can be performed. This process removed some unnecessary noise and silences, which decreased the final duration of the training set to $1382$ hours, the validation set to $8.3$ hours and the test set to $8.87$ hours.

Additionally, to test the effect of curriculum learning on different corpus sizes, we have provided results for models trained on: (a) $138$ hours ($10\%$ of the data), (b) $414$ hours ($30\%$ of the data) and (c) the whole data of $1382$ hours. For the 138h subset, we are training for 75 epochs, while for the 414h and 1382h subsets we train for 50 and 15 epochs, respectively.

\section{Experiments}

This work has been implemented as a python library on top of SpeechBrain  \cite{ravanelli2021speechbrain}
. Our models use logarithmic mel-based filterbanks as inputs and subwords as output units. Specifically, the transcriptions are tokenized using BPE \cite{bpe_article} with $1750$ output tokens.

Our architecture consists of a CRDNN encoder (CNN, RNN and DNN) and an RNN with attention as the decoder. Interested readers can find more details about the underlying implementation of these networks in SpeechBrain's documentation \cite{ravanelli2021speechbrain} and our github repository \textsuperscript{1}.  


The loss function used for training is the \textit{negative log likelihood}, which we will refer to as seq2seq loss. We also minimize the CTC loss on the output of the encoder for the first 10 epochs in the case of the full dataset and the first 15 epochs, otherwise. 

All of our models have been trained on a Nvidia Tesla A100 GPU card. To allow the processing of a bigger batch size, we used gradient accumulation which
aggregates the gradients and makes the update only after a certain number of steps (four in our case). This allowed us to use a batch size of 32.



\subsection{Training Strategies}

In this paper, we will provide an analysis of the following training strategies:
\textit{Duration-based} \textbf{(Baseline)}, \textit{Random Shuffling} \textbf{(RS)}, the training set is randomly shuffled before training. \textit{Seq2Seq Scoring} \textbf{(S2S)} is a loss-based scoring function using the seq2seq loss, while \textit{WER Scoring} \textbf{(WS)} uses a metric-based scoring function (using WER). \textit{WER Scoring w/ uniform Mixing} \textbf{(WS-M)} combines the metric-based scoring function with uniform mixing, and \textit{Seq2Seq Scoring w/ uniform Mixing} \textbf{(S2S-M)} is a loss-based strategy with uniform mixing. Lastly, \textit{Transfer Learning w/ WS-M} \textbf{(T-WS-M)} relies on the WER metric of an external model along with uniform mixing,while \textbf{(T-S2S-M)} uses the loss of the teacher model.
%
%
%
In the transfer-learning CL experiments, the teacher was chosen to be the baseline model trained on $1382$ hours of training data for $15$ epochs. 

\subsection{Results}

In Table \ref{table:cl_valid_testresults}, we include the results of models trained on each subset of the training data. To ensure that random initialization does not taint our experiments, we chose to train three separate models for each strategy and reported the averaged performance.

In all categories, our \textbf{WS-M} strategy has outperformed all of its competitors on both validation and test sets. Another interesting fact is that the \textbf{RS} strategy has also managed to consistently prevail over the baseline. Using Matched Pairs Sentence-Segment Word Error test, we observed no significant differences ($p<0.001$) among the three models trained with the \textbf{T-WS-M} or the \textbf{WS-M} strategy. On the other hand, we noticed that both the \textbf{RS} and the baseline strategy led to models where at least one out of the three runs performed significantly different from the rest. In addition, the standard deviation among the three runs of the baseline is $0.53$ compared to $0.45$ and $0.37$ for the \textbf{RS} and \textbf{WS-M} strategies, respectively. These findings suggest that the metric based scoring functions have a stabilizing effect which mitigates the impact of the random initialization.

\begin{table}[t!]
    \caption{Results of various CL strategies.}
    \label{table:cl_valid_testresults}
    \begin{center}
    \begin{tabular}{lcccc}
        \toprule
        
        \multirow{2}*{\textbf{Strategy}} & \multicolumn{2}{c}{\textbf{Valid}} & \multicolumn{2}{c}{\textbf{Test}} \\
         & \textbf{CER}  & \textbf{WER} & \textbf{CER} & \textbf{WER} \\
        \toprule
        \multicolumn{1}{c}{\textbf{138h Set}} &  &  &  &  \\
        Baseline & 14.03 & 38.47 & 16.14 & 43.56 \\
        RS & 13.15 & 37.55 & 16.15 & 43.69 \\ 
        \hdashline
        WS & 13.58 & 38.60 & 16.4 & 43.92 \\ 
        WS-M & \textbf{12.61} & \textbf{36.46} & \textbf{15.86} & \textbf{43.17} \\ 
        S2S-M & 14.45 & 40.48 & 19.21 & 50.13 \\ 
        T-WS-M & 13.55 & 37.97 & 17.3 & 46.07 \\ 
        \midrule
        \multicolumn{1}{c}{\textbf{414h Set}} &  &  & & \\
        Baseline & 13.19 & 32.57 & 13.61 & 35.88 \\
        RS & 9.31 & 28.43 & 12.48 & 34.74 \\ 
        \hdashline
        WS & 13.55 & 33.29 & 16.36 & 39.44 \\ 
        WS-M & \textbf{9.24} & \textbf{28.09} & \textbf{12.18} & \textbf{34.19} \\ 
        S2S-M & 10.39 & 30.97 & 13.17 & 37.03 \\ 
        T-WS-M & 9.76 & 29.47 & 12.86 & 35.65 \\ 
        \midrule
        \multicolumn{1}{c}{\textbf{1382h Set}} &  &  & & \\
        Baseline & 10.21 & 26.63 & 10.77 & 30.69 \\
        RS & 7.8 & 24.45 & 10.54 & 29.73 \\ 
        \hdashline
        WS & 22.01 & 47.17 & 23.33 & 50.95 \\ 
        WS-M & \textbf{7.47} & \textbf{23.27} & \textbf{10.14} & \textbf{28.98} \\ 
        S2S-M & 9.03 & 27.25 & 11.5 & 32.74 \\ 
        T-S2S-M & 8.04 & 24.85 & 11.19 & 30.82 \\ 
        T-WS-M & 7.98 & 24.7 & 11.1 & 30.88 \\ 
        \bottomrule
    \end{tabular}
    \end{center}
\end{table}

\begin{figure}[t]
  \centering
  \includegraphics[width=\linewidth,height=50mm]{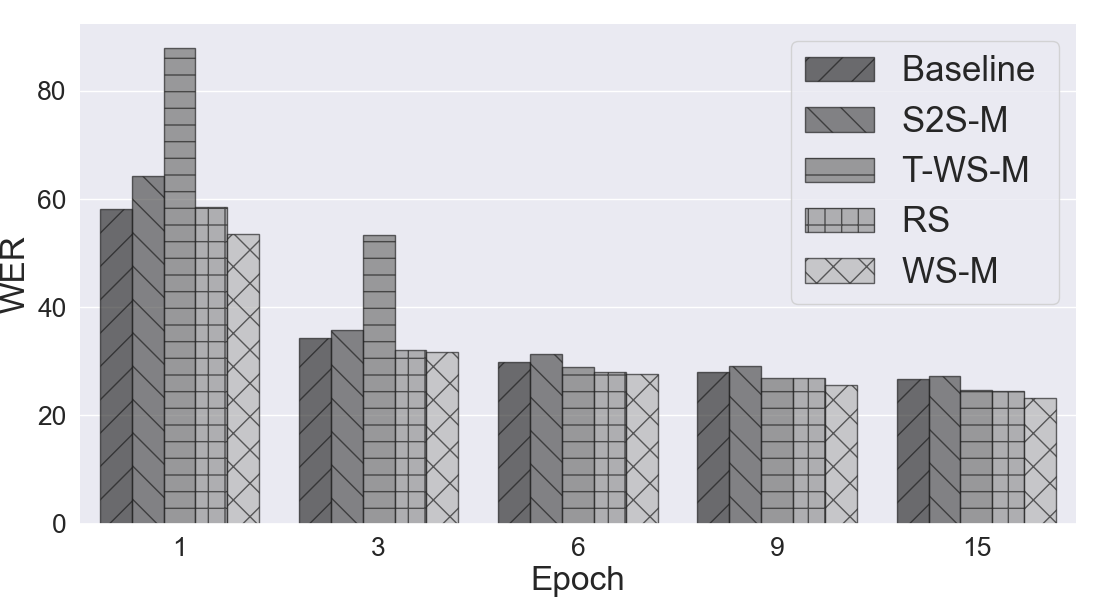}
  \caption{Evolution of the validation WER scores for the most representative models trained on the full training set.}
  \label{fig:baplot_full}
\end{figure}

Figure \ref{fig:baplot_full} shows the evolution of the WER on the validation set for 5 of our models. The performances of the \textbf{WS-M} and \textbf{RS} strategies are very close to each other and they both outperform the baseline at every epoch. 

\blfootnote{\textsuperscript{1}For further details see our github repository: https://www.github.com/aalto-speech/speechbrain-cl.}



\subsection{Pacing Function}

Table \ref{table:pacing_results} provides the results for the combination of the pacing function mentioned in \ref{pacing_func} and the \textbf{WS-M} and \textbf{T-WS-M} strategies, on the full training set of 1382 hours.
%
The column \textit{"Hours Seen"} refers to the amount of hours each model saw during the whole training cycle  in thousands. Even though the models trained with a pacing function, have only seen about half the training data, their performance is still on par with the rest of the models. The \textit{"Paced"} models have seen roughly the same amount of data as the models trained on the 138h train set, but they perform much better, due to the fact that they have seen different kinds of utterances and learned from them. This goes on to show the importance of curriculum in training deep neural networks.

\begin{table}[t!]
    \caption{Average WERs using a pacing function.}
    \label{table:pacing_results}
    \begin{center}
    \begin{tabular}{lccc}
        \toprule
        \textbf{Strategy} & \textbf{Hours Seen} & \textbf{Valid} & \textbf{Test} \\
        \toprule
        Baseline (138h) & \textbf{10.35K} & 38.47 & 43.56 \\
        Baseline (414h) & 20.70K & 32.57 & 35.88 \\
        Baseline (1382h) & 20.73K & 26.63 & \textbf{30.69} \\
        \hdashline
        (Paced) WS-M & 11.54K & \textbf{25.59} & 30.78 \\ 
        (Paced) T-WS-M & 11.82K & 26.2 & 31.79 \\ 
        \bottomrule
    \end{tabular}
    \end{center}
\end{table}

\section{Analysis}

\begin{figure}[t!]
  \centering
  \includegraphics[width=\linewidth]{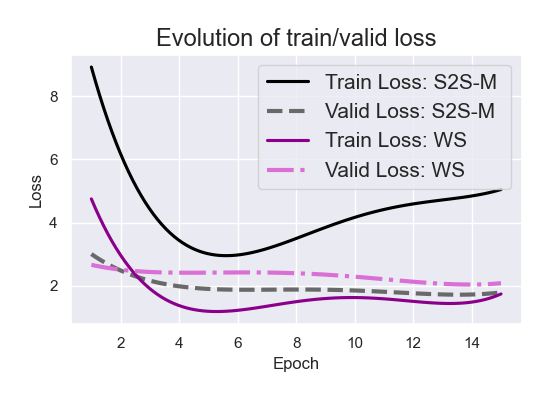}
  \caption{Underfitting and overfitting of the \textbf{S2S-M} and \textbf{WS} training strategies, for the 1382h train set.}
  \label{fig:loss_fits}
\end{figure}

The results in Table \ref{table:cl_valid_testresults} and Figure \ref{fig:baplot_full} expose some vulnerabilities of the \textbf{S2S-M}, \textbf{WS} strategies and the models trained with the \textit{Transfer-learning CL approach}.In particular, the \textbf{S2S-M} strategy has consistently underfitted the data while the \textbf{WS} strategy overfitted. These two observations are backed up in Figure \ref{fig:loss_fits} where the training loss stops improving rather quickly for the \textbf{S2S-M} strategy, but steadily improves for the \textbf{WS} in expense of the validation loss. The same progress was also followed by the respective models trained on $138$ and $414$ hours.

In the case of the \textbf{WS} strategy, we can attribute the overfitting to the fact that the corresponding model is only focusing on the easy examples, making it less able to generalize on the harder ones that always occur at the end of the epoch. This is also backed up by the fact that the \textbf{WS-M} strategy produces the best results in all cases, even though the underlying scoring function is the same. 

With regard to the \textbf{T-WS-M} strategy, Figure \ref{fig:baplot_full} makes it clear that the corresponding model is learning very slowly for the first 4 epochs and then suddenly improves.  The same behavior is also observed in the \textbf{T-S2S-M} strategy and the reason is that the transferred difficulty scores do not match the initial bad networks, and they need to achieve a certain level of performance before the curriculum makes sense.

\subsection{Training Time Analysis} \label{time_analysis}

The main motivation behind using the duration-based scoring function is that it minimizes padding and hence makes training faster \cite{ravanelli2021speechbrain}. In our experiments, we noticed that this is indeed the case. Table \ref{table:training_times} shows the overhead of the best training strategies, with regard to their total training time (amount of hours needed to train a whole model).

\begin{table}[t!]
    \caption{Training time overhead of our best CL strategies, compared to the baseline model. This also includes the time required to compute the curriculum scores.}
    \label{table:training_times}
    \begin{center}
    \begin{tabular}{lc}
        \toprule
        \textbf{Strategy} & \textbf{Overhead (\%)} \\
        \toprule
        RS & 15\% \\
        WS-M & 15\% \\
        T-WS-M & 4\% \\
        T-S2S-M & 0.6\% \\
        \hdashline
        (Paced) WS-M & -8\% \\ 
        (Paced) T-WS-M & -41\% \\ 
        \bottomrule
    \end{tabular}
    \end{center}
\end{table}

The random shuffling of the data causes a 15\% increase on the amount of time needed to train the model. The same also applies for the metric-based scoring function (with or without uniform mixing). On the other hand, when using the transfer learning CL approach, no significant change can be observed in the runtimes, making it a potential alternative to the duration-based metadata approach. Lastly, by using a pacing function, we get a marked decrease in the training time required while achieving a rather similar performance, compared to the baseline.

\section{Conclusion}

In this work, we proposed some novel techniques to apply curriculum learning in the training of e2e ASR systems. We saw that the \textit{metric-based with uniform mixing} scoring function consistently led to the best results, at the cost of slower training times. Additionally, through the use of a pacing function and the transfer-learning CL approach we reached a performance almost equal to the baseline, while training for about $60\%$ of the time.

Based on our findings, we believe that curriculum learning has a lot of potential in ASR. As future work, we plan to define new scoring and pacing functions that incorporate more information from the neural network under training. In addition, we are planning to test our approaches on some commonly used, large English datasets, to prove the effectiveness of our methods at a larger scale.



\section{Acknowledgements}
The computational resources were provided by Aalto ScienceIT.
We are grateful for the Academy of Finland project funding number 345790 in ICT 2023 programme's project "Understanding speech and scene with ears and eyes"





\clearpage

\bibliographystyle{IEEEtran}

\bibliography{curriculum}


\end{document}